\documentclass[preprint2, 10pt]{aastex}

\newcommand{\ex}[1]{\mbox{$\times 10^{#1}$}}

\newcommand{\kms}{\mbox{km s$^{-1}$}}
\newcommand{\apxx}{\mbox{$\alpha_{0.07}^{0.33}$}}
\newcommand{\appc}{\mbox{$\alpha_{0.33}^{4.9}$}}
\newcommand{\Jb}{\mbox{Jy bm$^{-1}$}}
\newcommand{\pyear}{\mbox{\% yr$^{-1}$}}
\newcommand{\Wmsr}[1]{\mbox{$\, \times ~10^{#1}$} {\mbox W~m$^{-2}$ Hz$^{-1}$ sr$^{-1}$}}

\shortauthors{Bietenholz et al.}
\shorttitle{Radio Spectral Index and Expansion of 3C58}

\begin{document}
      
\title{The Radio Spectral Index and Expansion of 3C58}

\author{M. F. Bietenholz}
\affil{Department of Physics and Astronomy, York University, Toronto, M3J~1P3, Ontario, Canada}
 
\author{N. E. Kassim}
\affil{Code 7213, Naval Research Laboratory, Washington, DC 20375-5320}
 
\author{K. W. Weiler}
\affil{Code 7213, Naval Research Laboratory, Washington, DC 20375-5320}
 
\begin{abstract}
We present new observations of the plerionic supernova remnant 3C58
with the VLA at 74 and 327 MHz.  In addition, we re-reduced earlier
observations at 1.4 and 4.9~GHz taken in 1973 and 1984.  Comparing
these various images, we find that: 1.  the remnant has a flat and relatively
uniform spectral index distribution, 2. any expansion of the remnant
with time is significantly less than that expected for uniform,
undecelerated expansion since the generally accepted explosion date in
1181~A.D., and 3. there is no evidence for a non-thermal synchrotron
emission shell generated by a supernova shock wave, with any such
emission having a surface brightness of $<1$\Wmsr{-21} at 327~MHz.
\end{abstract}

\keywords{ISM: individual (3C58) --- supernova remnants}

\section{INTRODUCTION}

Plerionic, or filled center supernova remnants (Weiler \& Panagia
1978; Weiler \& Sramek 1988) have
become established as a class.  The supernova remnant
\objectname[]{3C58} (\objectname[]{G130.7+3.1}) was an early member of
the class, and it has often been described as morphologically similar
to the Crab Nebula, the Crab being the prototype for this class.  3C58
is at a similar distance as the Crab, 3.2~kpc (Roberts et al.\ 1993),
has a similar size, $6' \times 9'$, and shows a similar, filamentary,
filled center morphology in the radio (Weiler \& Seielstad 1971;
Wilson \& Weiler 1976; Green 1986; Reynolds \& Aller 1988).  Like the
Crab, 3C58 has a relatively flat radio spectrum with $\alpha = -0.10
\pm 0.02 \; (S \propto \nu^{+\alpha})$ from 40~MHz to 15~GHz (Wilson
\& Weiler, 1976; Green 1986).  3C58 has been moderately well
identified with the historical supernova event of 1181~A.D. (Clark \&
Stephenson 1977), which would give it an age similar to that of the
Crab, which is associated with SN1054.

Plerions are thought to be powered by pulsars, which build the
non-thermal nebula through the injection of relativistic particles and
amplification of magnetic field, although the exact injection and
acceleration processes are not yet well understood.  In the optical,
radial velocities $< 900$~\kms\ have been measured for the filaments
in 3C58 (Fesen 1983), which is consistent with the association with
SN1181 only if the remnant has been substantially decelerated or the
filaments are due to the excitation of in-situ matter overrun by the
supernova shock.
However, the linear size of $<9$~pc (for a distance of 3.2~kpc) along
with the age of only 820~years imply rather low expansion velocities,
which suggests a low energy supernova event.
In addition, the radio flux density has been shown to be {\it
increasing}\/ (Aller \& Reynolds 1985; Aller, Aller, \& Reynolds 1986;
Green 1987).  It seems (Salvati et al.\ 1998; Woltjer et al.\ 1997)
that a non-standard evolution of the pulsar's output is required to
reconcile this and the relatively low frequency of the observed
spectral break in the synchrotron emission (Green \& Scheuer 1992)
with an age of 820 years.

\section{OBSERVATIONS AND DATA REDUCTION}

We observed 3C58 using the A and B configurations of NRAO's Very Large
Array (VLA) at 74~MHz using 63 spectral channels for a total bandwidth
of 1.5~MHz, and at 327~MHz using 31 spectral channels for a total
bandwidth of 3~MHz.  Spectral line mode was used to be able to excise
radio frequency interference (RFI).

The data were calibrated and reduced using NRAO's AIPS software
package.  The instrumental bandpass was determined using observations
of Cygnus~A.  The amplitude gains were set using Cygnus~A at 74,
and 3C48 at 327~MHz, and tied to the flux scale of Baars et~al.\
(1977).  Reduction of 74 MHz data requires somewhat different techniques
(see Kassim, Perley, \& Lazio 2000).  RFI is a significant problem
and the data were first automatically edited for deviant samples in
frequency space by the AIPS task SPFLG.  We then averaged the data
into a manageable number of channels and further edited each channel
manually.

Sidelobes from sources well outside the primary beam are a significant
problem at 74~MHz and Cassiopeia~A, which was 20\arcdeg\ away, produced
strong sidelobes in our field-of-view.  In order to correctly remove
these sidelobes, three-dimensional imaging is required (see Perley
1999) and the AIPS task IMAGR was used to simultaneously image and
deconvolve numerous flat but non-coplanar regions.  These were then
reassembled into a single image by the task FLATN.

\section{RESULTS}

\subsection{Images}

\begin{figure}
\plotone{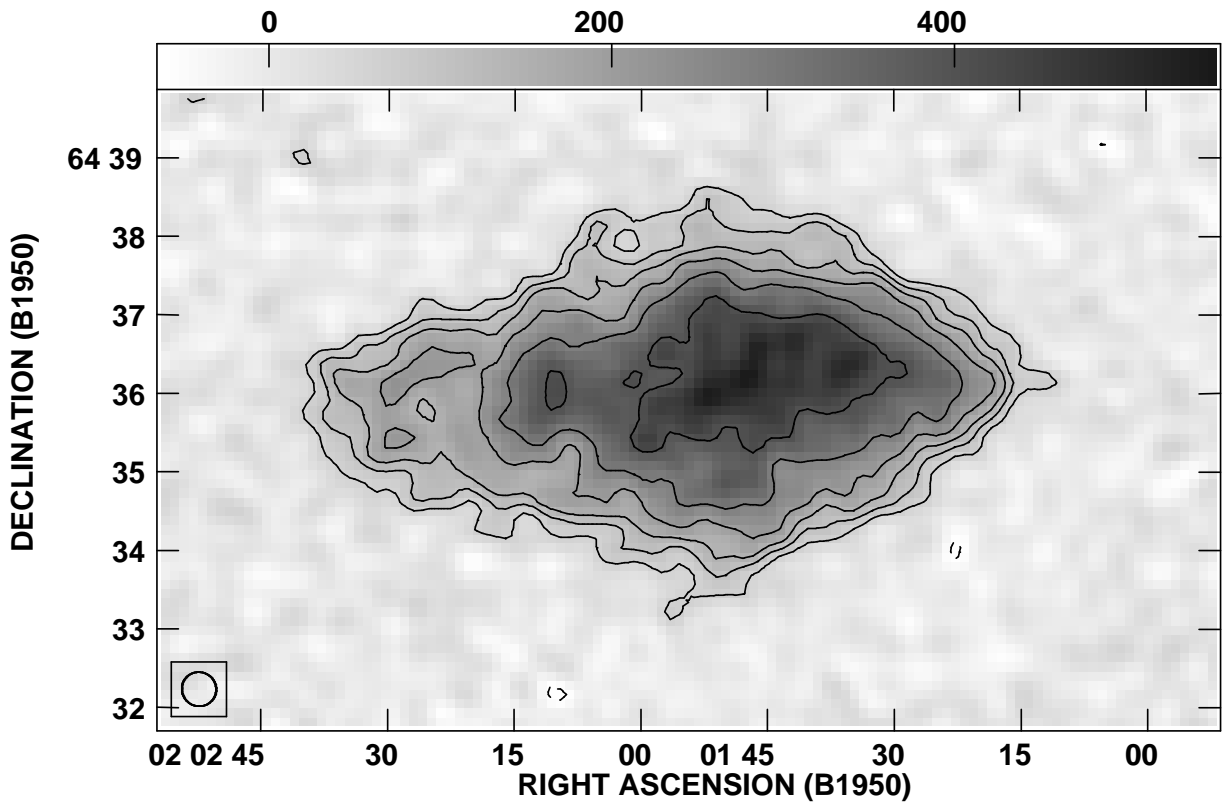} \figcaption[3c58-74map.ps]{Image of 3C58 at 74
MHz from combined VLA A \& B-array data.  The FWHM resolution was
26\arcsec\ and is indicated at lower left, the peak brightness was
0.55~\Jb, and the background rms was 20~m\Jb.  The greyscale is in
m\Jb. The contours are drawn at $-6$, 6, 10, 14.1, 20, 28.2, and 40\% of the
peak brightness.  \label{map4m}}
\end{figure}

\begin{figure}
\plotone{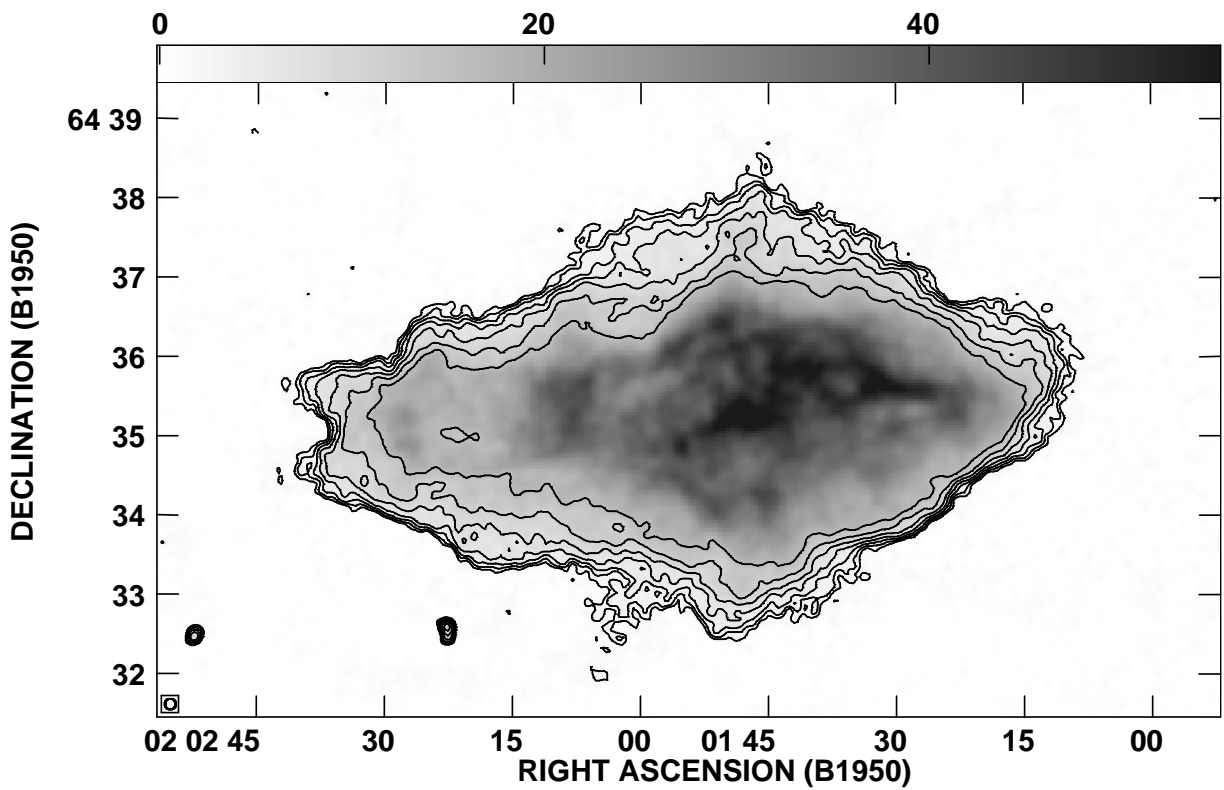} \figcaption[3c58-Pmap.ps]{Image of 3C58 at 327
MHz from VLA A and B-array data.  The FWHM resolution was 8.2\arcsec\
and is indicated at lower left, the peak brightness was 63~m\Jb, and
the background rms was 0.57~m\Jb.  The greyscale is in m\Jb. The
contours are drawn at $-2$, 2, 2.8, 4, 5.7, 8, 11.3 and 16~m\Jb.
\label{mapPsmall}}
\end{figure}

\begin{figure}
\plotone{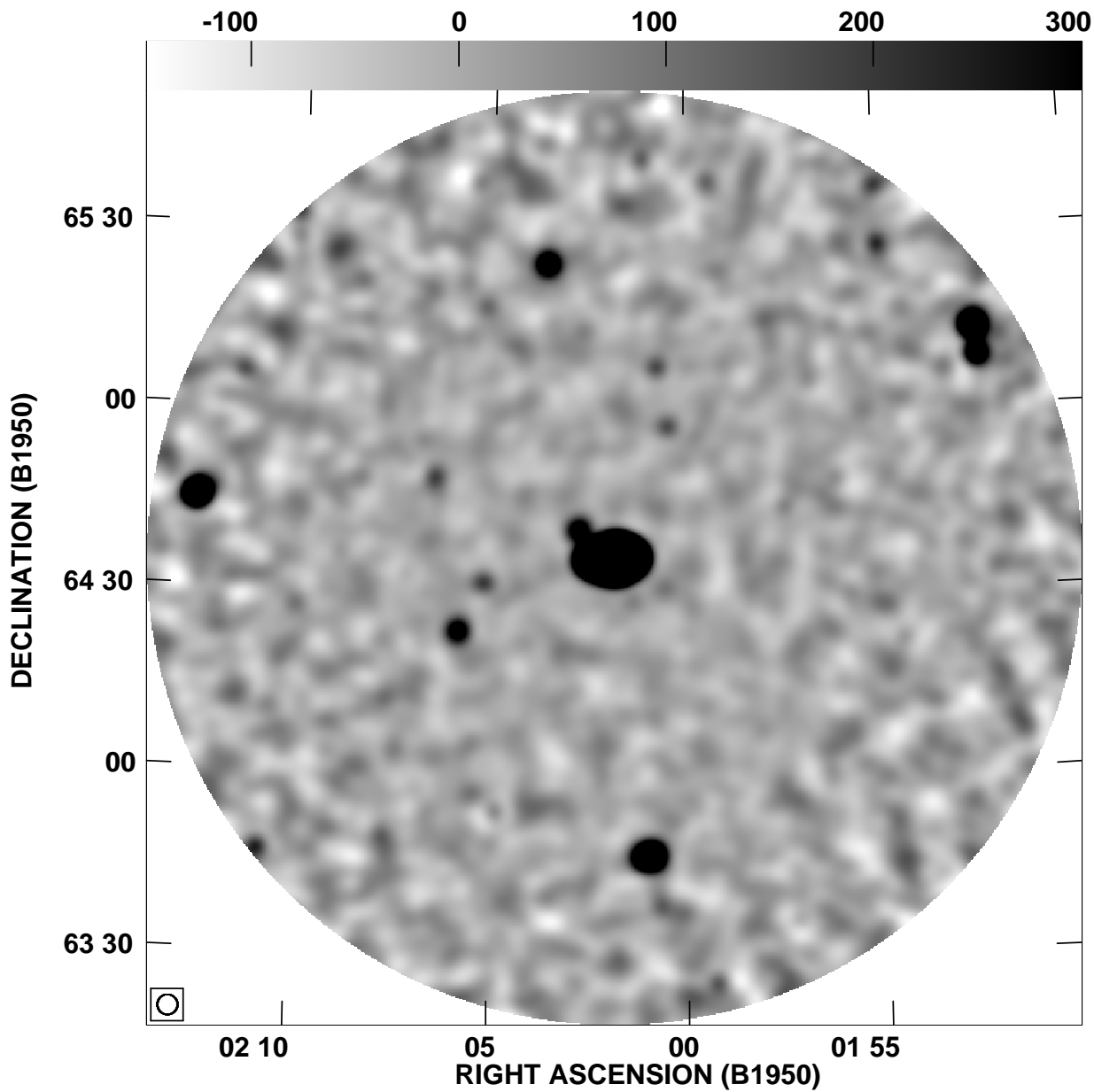}
\figcaption[3c58-big.ps]{A low-resolution image of the region
surrounding 3C58 at 327~MHz.  The image has been convolved with a
Gaussian of 3.3\arcmin\ FWHM, corrected for the primary beam response,
and blanked where the primary beam response is $<50\%$.  The peak flux
in the image was 15.9~\Jb. The background rms ranged from $\sim 20$~m\Jb\ 
near 3C58 to $\sim 45$~m\Jb\ near the edge of the field of
view.  The greyscale is in m\Jb. \label{mapPbig}}
\end{figure}

The full resolution image at 74~MHz is shown in Figure~\ref{map4m}.
This image was made from the combined A and B array data sets with
uniform weighting, and had a restoring beam size of 26\arcsec\ (FWHM).
The peak brightness for 3C58 was 0.55~\Jb, the background rms was
20~m\Jb, and the total flux density for 3C58 was 33.6~Jy.  The image
at 327~MHz is shown in Figure~\ref{mapPsmall}.  It was made from the
combined A and B array data sets, and had a restoring beam size of
8.2\arcsec\ (FWHM), a peak brightness for 3C58 of 63~m\Jb, a
background rms of 0.57~m\Jb, and a total flux density for 3C58 of
33.9~Jy.  It should be noted that no compact emission source is
detected near the center of the nebula.  In addition, in
Figure~\ref{mapPbig}, we show a low resolution image of the whole
primary beam area at 327~MHz, convolved to a resolution of 3.3\arcmin.
No shell emission is visible.  The average background rms between the
edge of the field and the center is 37~m\Jb, or 3.4\Wmsr{-22}.  We can
accordingly place a $3\sigma$ limit on any shell emission of
1\Wmsr{-21} at 327~MHz.  While the field of view at 74~MHz is even
larger, the sensitivity is considerably poorer, so the higher
frequency gives our most sensitive limit on extended emission.  We
note however, that the 74~MHz observations would be sensitive to very
large shells, i.e.\ with diameters $> 1.5\arcdeg$ and we see no such
shell emission.

\subsection{Spectral Index Distribution}

\begin{figure}
\plotone{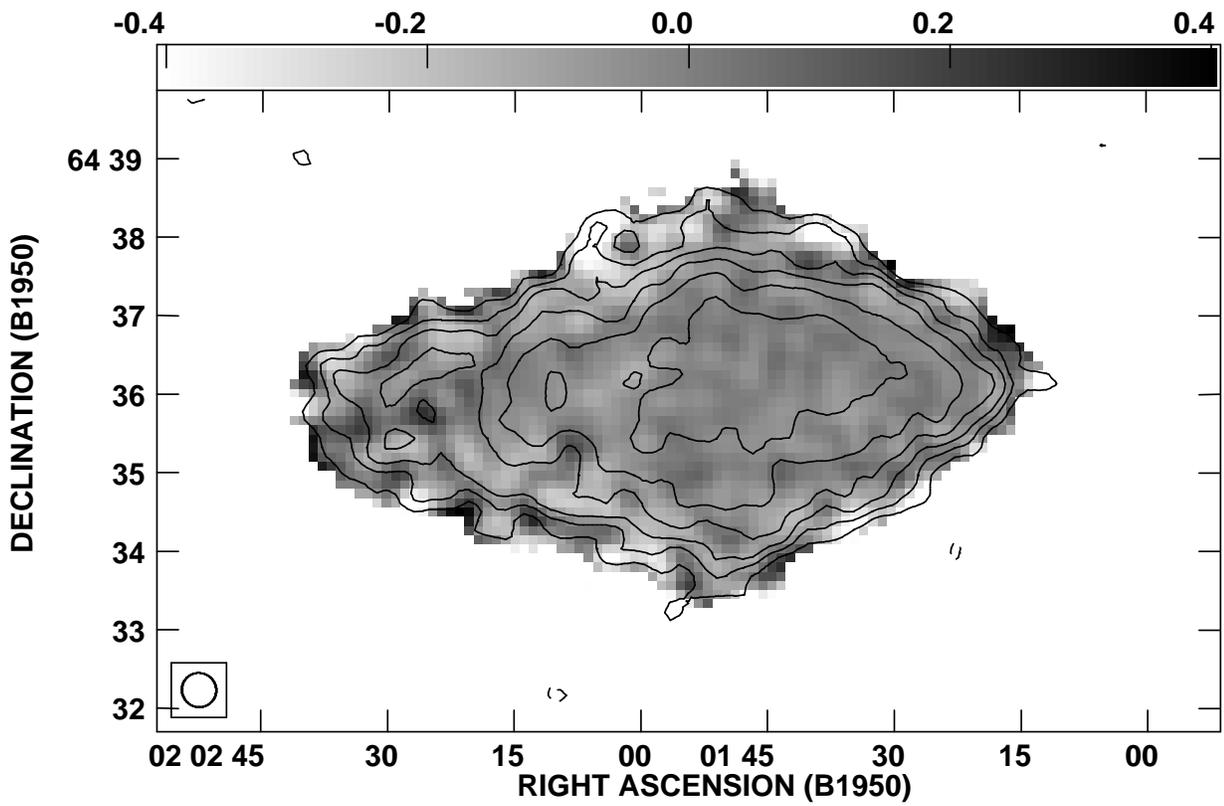}
\figcaption[3c58-4pspix.ps]{The radio spectral index of 3C58 between
74 and 327~MHz, \apxx, shown in greyscale.  The FWHM resolution is
21\arcsec\ (shown at lower left).  For reference, we repeat the total intensity
contours from Figure~\ref{map4m}.
\label{apxxmap}}
\end{figure}

We made an image of the spectral index between 74~MHz and
327~MHz\footnote{We will refer to spectral indices with sub- and
superscripts indicating the relevant frequencies in GHz.}, \apxx, as
follows.  The 327~MHz image was convolved to the larger resolution at
74~MHz and then the two images were aligned by the peaks of the point
sources near 3C58 (at 74~MHz, ionospheric refraction can cause
substantial displacement of the entire image).  This image of \apxx,
which has a resolution of 26\arcsec, is shown in
Figure~\ref{apxxmap} (regions where the uncertainty in \apxx\ was
larger than 0.4 have been blanked for clarity).  The spectral index
has a mean value of $-0.04$ over the nebula, with an rms of 0.18.
Only random variations are visible in Figure~\ref{apxxmap}.  In
particular, no steepening towards the edge of the nebula is observed.

We can calculate the integrated value of \apxx\ from the total flux
densities in the naturally weighted images, and we obtained
\apxx(integrated) = $-0.07 \pm 0.05$, where the uncertainty comes
chiefly from the uncertainty in the VLA amplitude calibration.  This
value is consistent with the $\alpha_{0.04}^{15} = -0.10 \pm 0.02$
given by Green (1986).

\begin{figure}
\plotone{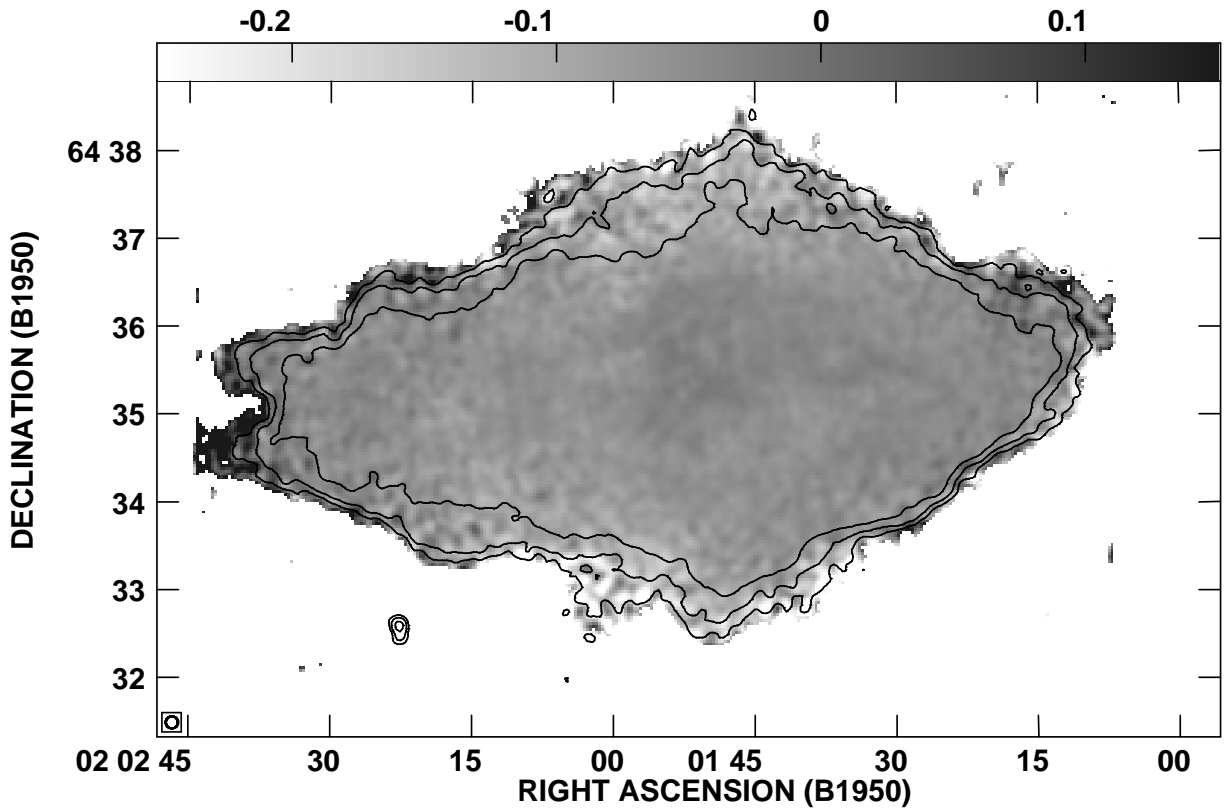}
\figcaption[3c58-pcspix.ps]{The radio spectral index of 3C58 between
327~MHz and 4.9~GHz, \appc, shown in greyscale.  The FWHM resolution
is 8.3\arcsec\ (shown at lower left).  For reference the 4, 8 and 16\%
contours from the total intensity image at 327~MHz are shown
also. \label{appcmap}}
\end{figure}

We also re-edited and re-imaged the data of Reynolds \& Aller (1988)
and made a spectral index image between 327~MHz and 4.9~GHz, \appc,
which is shown in Figure~\ref{appcmap}.  Again no particular features
are visible in the body of the nebula, and no steepening towards the
edge is observed.  The mean and rms scatter of the spectral index over
the central region of 3C58 were $-0.051$ and 0.014, respectively.
There is a very slight suggestion of a steepening of the spectral
index near the filaments by $<0.02$, which is likely not real, since
the systematic sources of error, in particular image fidelity limits
(see e.g., Perley 1999), will introduce errors in the spectral index
at this level which are likely correlated with the structure in the
image itself.  The integrated value of \appc\ was $-0.06 \pm 0.03$,
consistent with the above value from Green (1986).

\section{EXPANSION OF 3C58 \label{expsec}}

In order to determine the rate of expansion of the synchrotron nebula,
we compared our results to those from the aforementioned observations
in 1984 by Reynolds \& Aller (1988), and from observations in 1972 by
Wilson \& Weiler (1976).  We re-edited and re-imaged both these sets
of observations using MEM deconvolution.  In Table \ref{expimgs} we
give details of the images used for determining the expansion.  All of
the images at 1.4 and 5~GHz have been corrected for the attenuation of
the respective primary beam patterns (at lower frequencies the
correction is negligible).  Our goal was to determine the expansion
speed of the radio nebula.  Since there are few well-defined, compact
features, we choose to measure the expansion not by determining the
proper motion of individual features, but by determining an overall
statistical scaling between our images. We accomplished this by using
the MIRIAD task IMDIFF which calculates how to make one image most
closely resemble another by calculating unbiased estimators for the
scaling in size, $e$, the scaling and the offset in flux density, $A$
and $b$ respectively, and the offsets in RA and Dec, $x$ and $y$
respectively, needed to make the second image most closely resemble
the first.  We are chiefly interested in the expansion factor, $e$,
but because of uncertainties in flux calibration, absolute position,
and image zero-point offsets caused by missing short spacings, we
determined all parameters.  This method was developed by Tan \& Gull
(1985) and more details of its use in a similar situation are given by
Bietenholz et al.\ (1991).

Since we have shown above that the radio spectral index is uniform to
within the uncertainties, images at any two frequencies, $\nu_1$ and
$\nu_2$, should closely resemble one another, except for a flux
scaling factor which is given by $(\nu_1 / \nu_2)^\alpha$\/, where
$\alpha$ is the spectral index.  Thus we can usefully compare images
taken at different observing frequencies to test for expansion.

We ran IMDIFF on various combinations of the above images, always with
both images convolved to the same effective resolution, to get the
expansion factors shown in Table~\ref{expresults}.  In order to
estimate the uncertainty in our calculation of $e$, we also ran IMDIFF
on two pairs of (almost) contemporaneous images, for which $e$ should
be 0.0.  They are the expansion between 1.4 and 4.9~GHz at essentially
the same epoch, and the expansion between a CLEAN and a MEM image of
the 1973 4.9~GHz data, and the results are shown in the last two lines
in Table~\ref{expresults}.  In both these test cases we obtain
expansions of $<0.2$\%, which suggests that the uncertainties in our
expansion factors are of this order.

\begin{figure}
\plotone{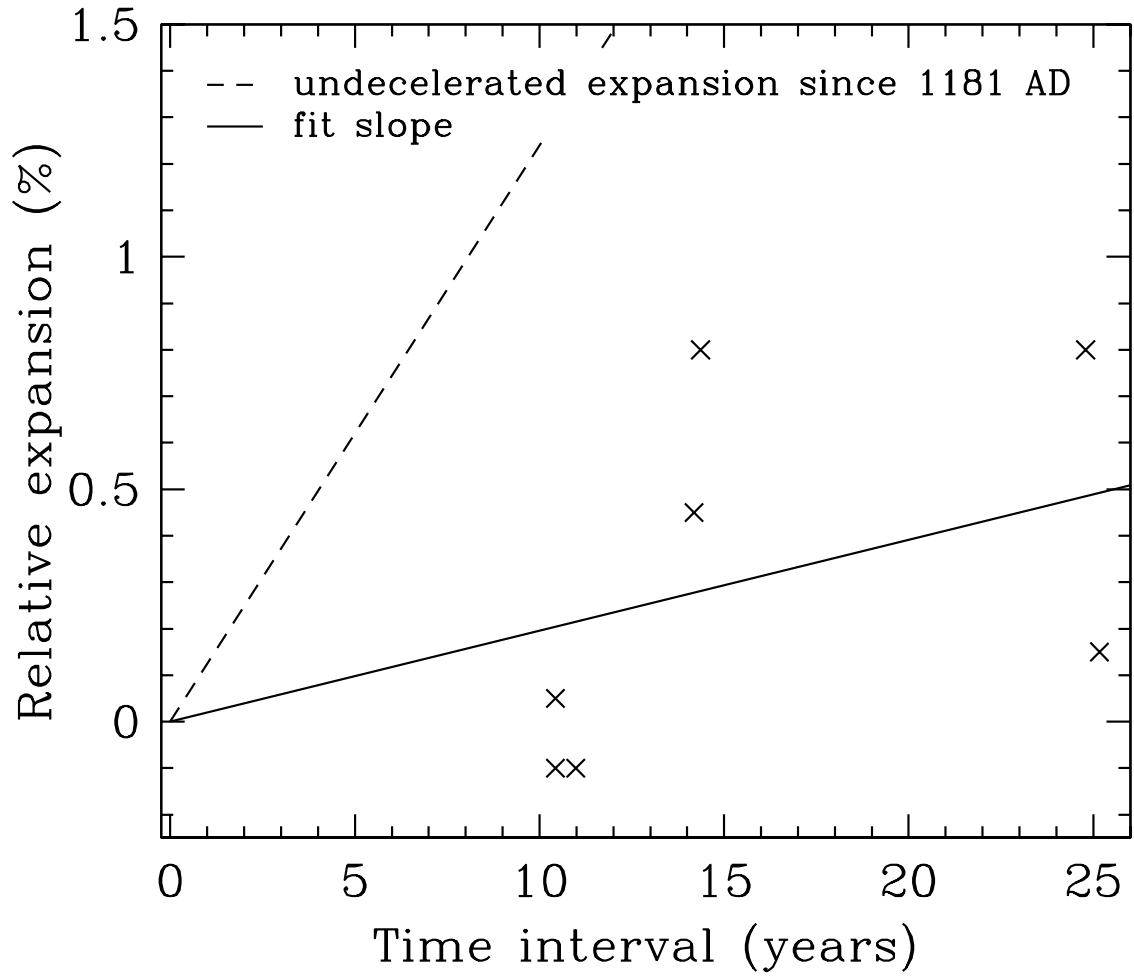} \figcaption[3c58expand8.ps]{The expansion of 3C58 over
various time intervals.  The points plotted represent the expansion
required by IMDIFF over the pairs of observations from
Table~\ref{expresults}, excluding the last two check points.  The
solid line indicates the fit expansion speed of $0.020 \; (\pm
0.08)$\pyear, and the dashed line indicates the expansion of
0.124\pyear\ expected for undecelerated expansion since
1181~A.D. \label{expplot}}
\end{figure}

In order to determine the expansion speed, we fit a straight line
through the first seven expansion factors in Table~\ref{expresults} by
weighted least squares.  The expansion for zero time difference must
be 0.0, so we constrain the fit line to pass through the origin.
Since we do not have formal uncertainties in the expansion factor, we
use the weights listed in Table~\ref{expresults}, which we derived as
follows: it seems reasonable to suppose that the uncertainty in the
expansion factor is proportional to the rms of the residuals to the
IMDIFF fit, $\sigma_{\rm fit}$, divided by the average brightness in
the images -- more flux in the image will allow a relatively better
fit for the same rms.  Since the total flux density is roughly the
same in all the images, the average brightness will be inversely
proportional to the number of beam areas over the source.  It also
seems reasonable that the uncertainty is inversely proportional to the
square root of the number of beam areas in the fitting region, $A_{\rm
fitting~region}$.  We thus adopt as the weight
\begin{equation}
W = \frac{1}{\sigma_{\rm fit}^2 \, A_{\rm fitting~region}}
\end{equation}
For the 5~GHz, 1973 WSRT data we use only expansions derived from a
version of the image high-pass filtered with a Gaussian of
1\arcmin~~FWHM, because this image had evident short-spacing problems
(Wilson \& Weiler 1976).  The points, as well as the weighted least
squares fit, are plotted in Figure~\ref{expplot}.

The straight line fit to the data in Table~\ref{expresults} by weighted
least squares is plotted as the {\it solid line} in
Fig.~\ref{expplot}.  The slope, or average expansion speed, we obtain
from the fit is of $0.020 \pm 0.008$\pyear, which
corresponds to $\sim 910 \pm 360$ and $550 \pm 220$~\kms\ along the
major and minor axes of the nebula, respectively.  We can estimate the
uncertainties from the scatter of the data points about the fit line,
and we find that a $W = 1$ corresponds to an uncertainty of
0.34\%\ in the expansion factor.  Our two check points, which measure
the expansion between almost simultaneous images (the last two lines
of Table~\ref{expresults}), are indeed consistent with an expansion of
0.0, as they should be.  We also note that an unweighted fit gives
almost the same expansion speed and uncertainty.

\begin{figure}
\epsscale{0.90} \plotone{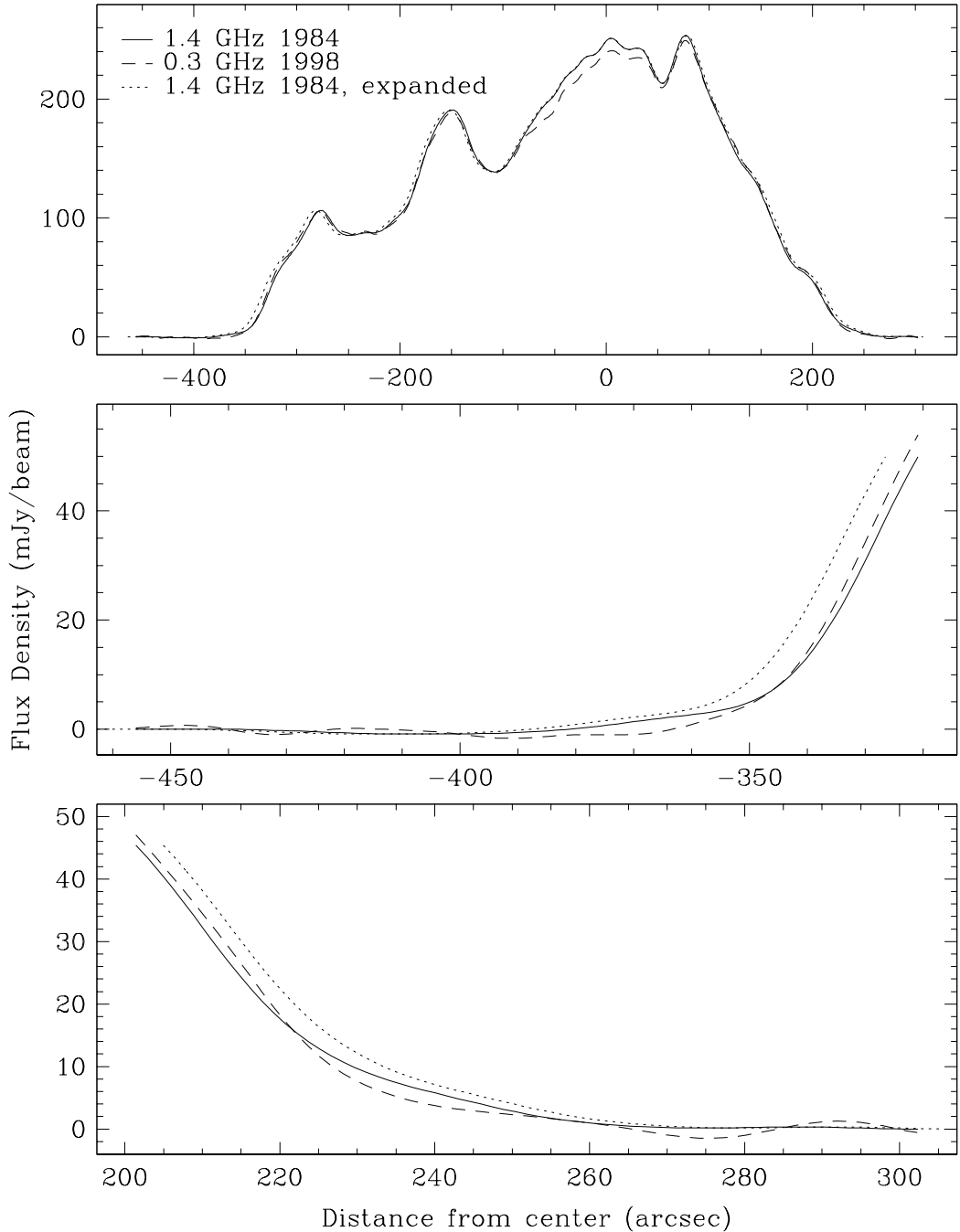}
\figcaption[3c58-3slice.ps]{Profiles through 3C58.  The top panel
shows the profile through 3C58, drawn at p.a.\ $-86$\arcdeg, ie.\
through the long axis of the nebula.  The solid line represents the
flux density at 1.4~GHz in 1984. The dashed line represents the
flux density at 0.3~GHz in 1998, multiplied by a scaling factor so as
to most closely match the 1984 profile (the scaling factor was
determined by IMDIFF, see text \S~\ref{expsec}).  The dotted line,
drawn for comparison, is the 1984 data homologously expanded by
1.74\%, which is the expected amount if 3C58 is the undecelerated
remnant of SN1181.  The middle and the lower panel are magnified
regions of the profile in the top panel, showing details of the
eastern and western limbs of 3C58 respectively. The FWHM resolution
was $17.7\arcsec \times 17.4\arcsec$ $@$ $-7.5\arcdeg$.\label{slicefig}}
\end{figure}

Our measured expansion speed ({\it solid line} in Fig.~\ref{expplot})
is much less than would be expected for undecelerated expansion of
3C58 since 1181~A.D.  The undecelerated ``historical'' expansion speed
would be 0.124\pyear\ and is plotted as the {\it dashed line} in
Fig.~\ref{expplot}.  In addition, in Figure~\ref{slicefig}, we show
profiles drawn through the long axis of 3C58 at 1.4~GHz in 1984 and at
0.3~GHz in 1998, and for comparison a hypothetical profile from
undecelerated expansion of the 1984 profile to 1998.  This last can be
seen to be notably larger than the true 1998 profile, also indicating
a lower than ``historical'' expansion speed.

The nebular structure is dominated by the relatively sharp outer edges
rather than by any well defined filaments or clumps, so our expansion
results derive in considerable part from the overall extent of the
nebula.  Unfortunately, this makes our expansion results sensitive to
short-spacing problems and also to the accuracy of the beam
corrections, especially at 5~GHz, where the VLA and WSRT primary beams
are 825 and 650\arcsec~FWHM, respectively.  Because of this we cannot
exclude the presence of systematic biases in our results.  However, we
feel that the aggregate expansion speed, which is derived from a
variety of combinations of different epochs, resolutions, telescopes,
and frequencies is unlikely to be greatly biased.

\section{DISCUSSION}

We measure an expansion rate of the radio nebula for 3C58 of $0.020
\pm 0.008$\pyear\ (corresponding to $\sim 910 \pm 360$ and $550 \pm
220$~\kms\ along the major and minor axes of the nebula,
respectively).  This is much smaller than would be expected if 3C58
were indeed the undecelerated remnant of SN1181, which would require
an expansion of 0.124\pyear\ ($\sim 5500$ and $\sim 3300$~\kms\ along
the major and minor axes respectively).  Conversely, our measured
expansion rate suggests an age for 3C58 of 5000~yrs with a $3\sigma$
lower limit of 2250~yrs, which is incompatible with the generally
accepted association with SN1181.  Although, we cannot entirely rule
out systematic sources of error in our determination of the expansion,
we think it unlikely that they are large enough to make our results
compatible with an expansion age of $\sim 820$ years.

This low expansion speed is corroborated by the proper motions of
optical filaments: Fesen, Kirshner, \& Becker (1988) and van den Bergh
(1990) looked for such proper motions and failed to find any larger
than 0.07\arcsec~yr$^{-1}$.  Fesen (1983) measured optical radial
velocities for several filaments in 3C58.  The largest radial
velocities he measured were $\sim$ 900~\kms, for filaments situated
near the projected center of the nebula, where the expansion is along
the line of sight, and thus where the expansion speed is expected to
be approximately equal to the radial velocity.  This is also
substantially lower than the $\gtrsim 3000$~\kms\ implied by
undecelerated expansion and a distance of 3.2~kpc. (We note that
Fesen, who used a somewhat smaller distance to 3C58 of 2.6~kpc,
reconciled these radial velocities with an age of 820~years by
suggesting that substantial deceleration had taken place, and that the
average speed since 1181~A.D. was $\sim 3000$~\kms).  In order to
determine an expansion age from this radial velocity, we need to know
the radius at which the measured filament lies from the center of the
nebula. If we take this radius to be the mean projected radius of
$\sim 4$~pc, these radial velocities also imply a large age of $\sim
4400$~years, but since the optical emission is faint, those radial
velocities may have been measured for filaments nearer the center of
the nebula, and thus a smaller radius and consequently a smaller age is
possible.  However, we consider it unlikely that the largest radial
velocities were measured for filaments only $\sim 1$~pc from the
center of the nebula. If 3C58 is associated with SN1181, all this
evidence suggests that it must have been greatly decelerated.

The simplest solution to these inconsistencies is to suggest that 3C58
is not the remnant of SN1181.  This also alleviates the difficulties
in accounting for the very sharp and rather low frequency spectral
break observed in 3C58 (Woltjer et al.\ 1997; Green \& Scheuer 1992).
Note that a lower value for the somewhat uncertain distance estimate
for 3C58 would reduce the age determined from the radial velocities,
but not that determined from the proper motions, so that is not a
solution.  The positional coincidence between SN1181 and 3C58,
however, was confirmed by a recent re-examination (Stephenson \&
Green 1999), and so if 3C58 is not to be associated with SN1181, it
must be asked where the remnant of SN1181 is, that there are no
other obvious candidates, and that it seems unlikely that the remnant
would already be too faint to observe at an age of only 820 years.

Van den Bergh (1990) suggests that the presently visible optical
filaments predate the supernova event, and are quasi-stationary but
energized by the passing supernova shock wave.  It is not clear that
the low expansion rate of the synchrotron bubble can be accounted for
in this way, since it seems unlikely that these pre-existing
filaments, which have a low filling factor, would be able to confine
the synchrotron bubble.  While some coupling between the synchrotron
bubble and the thermal filaments would be expected, since the magnetic
field in the bubble would wrap around the thermal filaments, one would
expect the synchrotron bubble to rapidly expand beyond the filaments
along with the rest of the supernova ejecta.  The outside edge of the
visible synchrotron bubble should be expanding at least at the speed
of the local ejecta, which would be $\sim 4000$~\kms\ for
undecelerated expansion.  Should pre-existing filaments nonetheless be
somehow confining the synchrotron bubble, then the filaments should be
accelerated.  The current measurements, unfortunately, do not have the
precision necessary to determine if this is the case.

A possible explanation for slow-moving filaments is the following.
The magnetic field will wrap around the pre-existing filaments, and
cause an intensification of the synchrotron emissivity in their
vicinity, which is why these filaments are visible in radio
synchrotron emission.  If the structure of the radio nebula is
dominated by these irregularities, then our determination of the
expansion would determine the expansion precisely of the filaments and
{\em not}\/ of the synchrotron bubble as a whole.  We note that
expansion speeds we measure at lower effective resolutions tend to be
slightly smaller than those at larger resolutions, as would be
expected in this case, but even the measured expansions at lower
resolutions are far below the expected values for undecelerated
expansion.  Furthermore, profiles through the images do not support
the conjecture of quasi-stationary filaments in an expanding bubble:
the profiles displayed in Figure~\ref{slicefig} show that even the
expansion of the edge of the nebula appears to be considerably less
than is expected from undecelerated expansion.

Let us turn to some general physical considerations about the
synchrotron bubble.  The pressure in the synchrotron bubble is
relatively high: standard arguments give a minimum pressure of
$>10^{-10}$~dyn~cm$^{-2}$. This is the minimum pressure of the
synchrotron fluid assuming a filling factor of 1.0, and a line of
sight depth equal to the long axis of the projected nebula.  Both a
lower filling factor and the admixture of any thermal material would
raise the pressure, and we think it unlikely that the line of sight
dimension is much larger than the long axis of the projected nebula.

The slow expansion of the bubble therefore
necessitates some confining pressure, since the sound and Alfv\'{e}n
velocities in the bubble are also high.  This leads to the question of
what surrounds the presently visible nebula.  The three possibilities
are a) the ISM b) the wind bubble of the pre-supernova progenitor of
SN1181, or c) a freely expanding stellar envelope thrown off in the
supernova explosion.  This last might be expected, since the total
energy in the visible remnant is far below the expected energy of
$10^{51}$~ergs released in a supernova for any reasonable mass in the
filaments.  This last situation does seem to obtain in the Crab
Nebula, where strong circumstantial evidence points to the existence
of such a freely expanding stellar envelope (Hester et al.\ 1996;
Sankrit \& Hester 1997), even though the radio emission from this
shell has so far eluded detection (Bietenholz et al.\ 1997; Frail et
al.\ 1995) and the direct optical detection is controversial.

The static pressure of any of these three exterior media, $p_{\rm ext}$
is expected to be considerably smaller than the minimum pressure in
the synchrotron bubble:
\begin{equation}
p_{\rm ext} = 1.4\ex{-12} {\rm ~dyn~cm}^{-2} \; [n / 1cm^{-3}] \; [T / 10^4 {\rm K}]
\end{equation}
where $n$ is the number density and $T$ is the temperature.
This suggests that the synchrotron bubble is confined by ram pressure
as it expands into the surrounding material.  We are then brought to a
crucial difference between 3C58 and the Crab Nebula: in the case of the
Crab, the synchrotron bubble has been accelerated since the supernova
event (Bietenholz et al.\ 1991), and is therefore expanding {\em
into}\/ the stellar envelope, and confined by the ram pressure of this
expansion.  In the case of 3C58, by contrast, the synchrotron bubble
is expanding considerably more {\em slowly}\/ than any freely
expanding material thrown off in SN1181, and cannot therefore be
confined by ram pressure in any freely expanding ejecta.

This implies for 3C58 that there is no freely expanding envelope,
making it a very low energy supernova event, or that the envelope has
been greatly decelerated, which seems unlikely in light of the
non-detection of any shell emission exterior to the synchrotron
remnant, or that it is not the remnant of SN1181.  The ram pressure
required suggests that the external medium have a number density,
$n_{\rm ext}$, of
\begin{equation}
   n_{\rm ext} \sim 0.1 {\rm ~cm}^{-3} \; [v_{\rm shock}/1000 \kms]^2 \; [m/m_{\rm H}]
\end{equation}
where $v_{\rm shock}$ is difference in velocity between the
synchrotron bubble and the undisturbed exterior medium, and $m$ is the
mean mass per particle.

A possible scenario would be that 3C58 is the composite remnant of a
binary where both stars have undergone supernova explosions, with the
earlier supernova leaving the slowly expanding remnant, and the later,
low energy one providing the historical supernova event in 1181~A.D\@.
This scenario seems rather contrived, since having two supernovae
within a few thousand years of one another is quite unlikely

We will turn to the non-detection of any shell emission.  If there
was an envelope ejected in SN1181, as discussed above, it must have
been decelerated by at least a factor of 3, which in turn suggests
that it has already swept up considerable external material, and is
well into the Sedov phase of its evolution.  It is then surprising that
the radio brightness of this exterior shell is considerably less then
of the faintest known shells.  Again, it is instructive to point out the
difference between 3C58 and the Crab Nebula.  Also for the Crab, radio
emission from the shell has not yet been detected to similar limits
in surface brightness, but in the case of the Crab, the envelope is
still in the free expansion phase, and is not expected to have
converted much of its energy into relativistic particles, and therefore
is expected to be fainter than most shell remnants.

Could a shell for 3C58 be masked because it is largely coincident with
the synchrotron nebula?  Shock accelerated radio-synchrotron emitting
electrons typically produce radio emission with a spectral index,
$\alpha, \; \sim -0.6$.  Our spectral index results, however, indicate
that there is no significant radio emission with $\alpha = -0.6$ near
the edge of the nebula, since we observe no steepening of the spectrum
there.  Also, the total flux density we measure at 74~MHz gives an
integrated spectral index no steeper than that at higher frequencies,
contrary to what would be expected if there were a steep spectrum
component present.

Both \apxx\ and \appc\ show no spatial variation over the nebula,
\appc\ showing rms variations of only $<0.05$.  The uniformity of the
spectral index suggests that a single acceleration mechanism is
responsible for all the radio emitting electrons.  This is thought to
be an as yet undetected pulsar, although there are a number of
problems with 3C58 in this regard.  Recent X-ray results by Bocchino
et al.\ (2001) do not show the expected black-body X-ray emission from
the neutron star.  No pulsed emission has so far been found at any
wavelength, although in a deep radio image, Frail \& Moffett (1993)
detected a ``wisp'' like feature, similar to ones usually associated
with the termination shock of a pulsar wind.

\section{A SHELL AROUND 3C58}

In case of the Crab Nebula, there is substantially less kinetic energy
in the remnant than the $10^{51}$~ergs expected from a type II
supernova.  This has suggested the existence of a rapidly moving shell
around the presently visible remnant which carries the excess energy
and also confines the relatively high-pressure synchrotron bubble.
Much observational effort has been expended in search of this shell
around the Crab, but no shell has yet been detected in the radio,
ruling out any radio emission with a surface brightness much less than
that of the faintest known young shell remnant, SN1006 (Frail et al.\
1995 and Bietenholz et al.\ 1997).  However, in the case of the Crab,
circumstantial evidence for the existence of expanding material
outside the presently visible nebula has been made from Hubble Space
Telescope observations (Hester et al.\ 1996) which show the
Rayleigh-Taylor unstable interface between the high-pressure
synchrotron bubble and the hydrogen shell.

In the case of 3C58, there is a similar deficit in energy, especially
in view of the low expansion velocity of both the synchrotron bubble
and the optical filaments, and so a rapidly moving shell might also be
expected exterior to the presently visible nebula.  Several searches
for a radio shell around 3C58 have been made, notably by Reynolds \&
Aller (1985), and no sign of radio emission from a shell has yet been
detected.  

Our 327~MHz results give a sensitive upper limit on the radio emission
from a putative shell of 1\Wmsr{-21}\ at that frequency.  Scaling with
an assumed shell spectral index of $-0.6$ this is equivalent to
4\Wmsr{-22}\ at 1.4~GHz, thus very slightly more sensitive than the
earlier limit of 4.7\Wmsr{-22}\ obtained by Reynolds \& Aller (1985).
We are, however, more sensitive to steep-spectrum shells and also to
larger shells due to our larger field-of-view at 327~MHz.  We note
that, even on our 74~MHz images, where the primary beam radius is
$\sim 5\arcdeg$, there is no sign of shell emission, albeit at a lower
sensitivity than at 327~MHz.  A sensitivity to larger shells may be
germane if 3C58 is, in fact, older than SN1181 since at 10,000~\kms, a
3500 year old shell at 3.2~kpc would have an angular diameter of
1.2\arcdeg.

\section{CONCLUSIONS}

New observations at 74 and 327 MHz and reanalysis of older
observations dating back to 1974 indicate that: 

\begin{trivlist}

\item{1.} 3C58 has a flat spectral index of $\apxx = -0.05$, with little
variation over the nebula.

\item{2.} The expansion of the synchrotron nebula is much slower than
expected for an explosion date in 1181~A.D. with constant,
undecelerated expansion since then. A possible explanation is that
3C58 is not the remnant of SN1181 but is far older with an age of
$\sim 5000$ years, however we are reluctant to abandon the generally
accepted association until a better understanding of the dynamics of
plerions is available and a better candidate remnant for SN1181 is
found. An alternate explanation is that 3C58 has been greatly decelerated.

\item{3.} We are also able to show that, as in the Crab Nebula, any
non-thermal emission a from a shock wave outside the observable
plerion has very low surface brightness of $<1$\Wmsr{-22}.  If 3C58
has been greatly decelerated as suggested in 2.\ above, then one would
expect radio emission from such a shock wave.

We will close by summarizing several apparent inconsistencies in the
observational material on 3C58.  First, the measured low proper
motions, both in the radio and in the optical, along with the low
synchrotron break frequency suggest an age much older than 820~years.
On the other hand, the positional association with SN1181 seems
secure, and no other remnant has been identified for this historical
supernova. The radio spectral index is very uniform, suggesting a
common source for the radio-emitting electrons, namely a pulsar wind,
yet the postulated pulsar has escaped detection and the expected X-ray
emission from the hot surface of the neutron star has not been
detected.  Finally, the presently observable remnant contains only a
small fraction of the kinetic energy expected from a type~II
supernova.  Unless SN1181 was an anomalously low energy event, a shell
consisting of the rapidly moving external layers of the star would be
expected to carry the balance of the kinetic energy.  Such a shell has
not been detected, and is anomalously faint in comparison to other
known shells.  Future study will hopefully allow us a more complete
and consistent picture of this fascinating object.

\end{trivlist}

\acknowledgements

The National Radio Astronomy Observatory is a facility of the National
Science Foundation operated under cooperative agreement by Associated
Universities, Inc.  Research at York University is partly supported by
NSERC\@.  NEK and KWW wish to thank the Office of Naval Research (ONR)
for the 6.1 funding supporting this research.  We thank Kristy Dyer
and Stephen Reynolds for making the 1984 data available to us, and
Mark Bentum for help with retrieving the 1973 WSRT data. We thank
the referee for several useful suggestions.

\onecolumn

\clearpage

\begin{deluxetable}{r@{ }c@{ }l l c c c r@{ $\times$ }r@{ $@$ }r l}
\tablecaption{Images used for the Expansion Determination \label{expimgs}}
\tablewidth{0pt}
\tablehead{
\multicolumn{3}{c}{Date} 
                         &\colhead{Telescope} &\colhead{Freq.}
                         &\colhead{$S_{total}$\tablenotemark{a}}
                         &\colhead{Image rms} 
                         &\multicolumn{3}{c}{Convolving Beam}   
                         &\colhead{Reference} \\
\colhead{} & \colhead{}  &\colhead{} & \colhead{}
                         &\colhead{(MHz)}    
                         &\colhead{(Jy)} &\colhead{(mJy/bm)}
                         &\colhead{(\arcsec} &\colhead{~\arcsec} &\colhead{~\arcdeg)}  \\
}
\startdata
\multicolumn{2}{c}{Feb.-Jun.} 
          & 1973  &  WSRT & 4995 & 30.1 & 2.44  &  8.3 & \multicolumn{2}{l}{8.3} &Wilson \& Weiler 1976 \\
%
 15 & Dec.& 1973  &  WSRT & 1415 & 33.1 & 1.35  & 22.7 & 20.9 &   0.0 &Wilson \& Weiler 1976 \\
%
\multicolumn{2}{c}{Jan.-Aug.} 
          & 1984  &  VLA  & 4866 & 29.5 & 0.43 & 8.3 & \multicolumn{2}{l}{8.3} &Reynolds \& Aller 1988 \\
\multicolumn{3}{c}{ '' } & '' &  ''  &  ''  & 1.71 & 17.7  & 17.4 &$-7.5$ & '' \\
%
\multicolumn{2}{c}{Jan.-Dec.}
          & 1984  &  VLA  & 1446 & 33.9 & 0.49 & 17.7  & 17.4 &$-7.5$ & Reynolds \& Aller 1988 \\
  1 & Oct.& 1998  &  VLA  & \phn327 & 33.0 & 1.17  & 17.7 & 17.4 & $-7.5$ & {\em this paper; B-array only}\\
\multicolumn{2}{c}{Mar.-Oct.} & 1998
                  &  VLA  & \phn327 & 34.0 & 0.58  &   8.3 & \multicolumn{2}{l}{8.3}  & {\em this paper; A+B arrays}\\
\enddata
\tablenotetext{a}{$S_{total}$\/ is the total cleaned flux density in the image.}
\end{deluxetable}

\clearpage

\begin{deluxetable}{r c@{\protect\phantom{xxxxx}} c c c c c c r}
\tablecaption{Expansion of 3C58\label{expresults}}
\tablewidth{0pt}
\tablecolumns{9}
\tablehead{
\multicolumn{2}{c}{First Image}  & \multicolumn{2}{c}{Second Image}
   & \colhead{Resolution\tablenotemark{a}}
   & \colhead{Time Interval}     & \colhead{Change in Size} 
   & \colhead{Rms of Fit}        & \colhead{Weight\tablenotemark{b}} \\
\colhead{Year\tablenotemark{c}} 
               & \colhead{Freq.} & \colhead{Year\tablenotemark{c}}  
               & \colhead{Freq.} \\
               & \colhead{(MHz)} &                 & \colhead{(MHz)}
               & \colhead{~~~(\arcsec)} & \colhead{ (years) } 
               & \colhead{~~~(\%) }     & \colhead{(mJy/bm)}
}
\startdata
 1984.39 & 1.4 & 1998.75 & 0.3 & 17 & 14.4 & \phs 0.80  & 1.4 & 1.1 \\
 1984.27 & 4.9 & 1998.46 & 0.3 &\phn 8 & 14.2 & \phs 0.45  & 0.6 & 1.7 \\ 
 1973.96 & 1.4 & 1998.75 & 0.3 & 21 & 24.8 & \phs 0.80  & 3.0 & 0.5 \\ 
 1973.96 & 1.4 & 1984.39 & 1.4 & 21 & 10.4 &    $-0.10$ & 2.2 & 0.9 \\ 
 1973.96\tablenotemark{d}& 1.4 & 1984.39\tablenotemark{d} 
                         & 1.4 & 21 & 10.4 & \phs 0.05  & 1.8 & 1.3 \\
 1973.29\tablenotemark{d}& 4.9 & 1998.46\tablenotemark{d} 
                         & 0.3 &\phn 8 & 25.2 & \phs 0.15  & 0.9 & 0.8 \\
 1973.29\tablenotemark{d}& 4.9 & 1984.27\tablenotemark{d}
                         & 4.9 &\phn 8 & 11.0 &    $-0.10$ & 0.7 & 1.3 \\
\sidehead{The results below are included as a check}
 1984.27 & 4.9 & 1984.39 & 1.4 & 17 &  0.1 & $-0.20$    & 1.4 & 1.4 \\ 
 1973.27\tablenotemark{e}& 4.9 & 1973.27\tablenotemark{e} 
                         & 4.9 & \phn 8 &  0.0 &\phs0.20     & 0.9 & 0.8 \\ 
\enddata
\tablenotetext{a}{The rounded average resolution, see Table~\ref{expimgs}.}
\tablenotetext{b}{These are the weights used for fitting the overall
expansion; see text for details.}
\tablenotetext{c}{For observations involving multiple arrays, the date given is the average over all the arrays, weighted by the length of the individual array observing runs.}
\tablenotetext{d}{These images have been high-pass filtered using
a circular Gaussian of FWHM 60\arcsec.}
\tablenotetext{e}{This line pertains to the difference between CLEAN and MEM images from 
the same data}
\end{deluxetable}

\end{document}